\documentclass[prl,twocolumn,superscriptaddress,footinbib,showpacs]{revtex4}

\usepackage{graphicx}
\usepackage{epsf}

\newcommand{\bq}{\begin{equation}}
\newcommand{\eequ}{\end{equation}}
\newcommand{\bqa}{\begin{eqnarray}}
\newcommand{\eqa}{\end{eqnarray}}

\def\kt{\ensuremath{\tilde k}}

\def\expct#1{\ensuremath{\langle #1\rangle}}

\def\Xe{\ensuremath{\expct{X}_\mathrm{e}}}
\def\Pe{\ensuremath{\expct{P}_\mathrm{e}}}

\def\Vxe{\ensuremath{{V_X^\mathrm{e}}}}
\def\Vpe{\ensuremath{{V_P^\mathrm{e}}}}

\def\Ce{\ensuremath{C_\mathrm{e}}}

\def\Heff{\ensuremath{H_\mathrm{eff}}}
\def\capsubsz{\scriptscriptstyle}

\def\omegaHO{\ensuremath{\omega_\mathrm{\capsubsz HO}}}
\def\prty{\ensuremath{\mathcal{P}}}

\def\Vmax{\ensuremath{{V_\mathrm{max}}}}

\begin{document}
\title{Quantum Feedback Control of Atomic Motion in an Optical Cavity  
                    \vbox to 0pt{\vss
                    \hbox to 0pt{\hskip-37pt\rm LA-UR-03-6826\hss}
                    \vskip 25pt}}

\author{Daniel A. Steck}
\affiliation{Theoretical Division (T-8), MS B285, Los Alamos National
  Laboratory, Los Alamos, NM 87545}
\author{Kurt Jacobs} 
\affiliation{Theoretical Division (T-8), MS B285, Los Alamos National
  Laboratory, Los Alamos, NM 87545}
\affiliation{Centre for Quantum Computer Technology, 
Centre for Quantum Dynamics, School of Science,
Griffith University, Nathan 4111, Australia}
\author{Hideo Mabuchi}
\affiliation{Norman Bridge Laboratory of Physics 12-33,
California Institute of Technology, Pasadena, CA 91125}
\author{Tanmoy Bhattacharya}
\affiliation{Theoretical Division (T-8), MS B285, Los Alamos National
  Laboratory, Los Alamos, NM 87545}
\author{Salman Habib} 
\affiliation{Theoretical Division (T-8), MS B285, Los Alamos National
  Laboratory, Los Alamos, NM 87545}

\begin{abstract}
We study quantum feedback cooling of atomic motion in an optical cavity.
We design a feedback algorithm that can 
cool the atom to the ground state of the optical potential 
with high efficiency despite the nonlinear nature of this problem.
An important ingredient is a simplified state-estimation algorithm, 
necessary for a real-time implementation of the feedback loop.
We also describe the critical role of parity dynamics in the
cooling process and  present a simple theory that 
predicts the achievable steady-state atomic energies.
\end{abstract}
\maketitle

The control of quantum systems is a problem that lies at the heart
of several fields, including atom optics and nanomechanics.
Many of the techniques developed thus far for controlling quantum
systems in an initially unknown state involve the use of dissipative 
processes to achieve a well-defined final state, as in the
laser cooling methods in atom optics \cite{Metcalf99}.
A different paradigm for cooling 
involves modifying the system Hamiltonian based on information 
provided by measurements (i.e., quantum feedback control).
Such a strategy for control is generally applicable to 
quantum as well as classical systems 
\cite{Raizen98, Belavkin99, Doherty00, Armen02, Vitali03, Hopkins03, Geremia03}.
As quantum-limited measurement techniques become more available,
quantum feedback control is likely to soon be
an essential element in atom-optical and nanomechanical toolkits.

One of the most promising avenues for the study of quantum
feedback control centers on experiments in cavity quantum electrodynamics (CQED), 
where a single quantum system may be monitored continuously
in real time, as has been demonstrated in a series of
pioneering experiments \cite{Hood98, Mabuchi99, Hood00, Pinkse00}.
Furthermore, the optical potential due to the cavity light
provides an easily adjustable means for applying forces on
the atom.  
Feedback has been shown to increase the storage time
for atoms in 
such an experimental system \cite{Fischer02}. 
Despite these efforts, however, CQED 
feedback cooling has not yet been conclusively demonstrated,
highlighting the need for a deeper theoretical understanding
of this system as well as more sophisticated, higher performance
cooling algorithms.
Our goal here is to develop approximate
estimation  and  control algorithms for cooling
the atomic motion in such a CQED system---an ``active cooling''
approach that is distinct from passive cavity cooling methods
\cite{Horak97}.  
We will then analyze the performance of these
algorithms both analytically and through numerical simulations.
This problem is particularly interesting due to the nonlinear
form of the measurement operator,
a situation distinct from previous studies of nonlinear quantum
feedback control \cite{Doherty00, Wiseman02}.
An important issue that we address here is whether the (Gaussian)
estimation methods developed for linear systems 
are still effective in this problem.

The setup we are considering is an atom 
in a microcavity in the
strong-coupling regime, as in the aforementioned CQED
experiments \cite{Hood98, Mabuchi99, Hood00, Pinkse00}, but
where the output light is monitored via
homodyne detection \cite{Wiseman93}, which gives information
about the atomic position in the cavity \cite{Quadt95}.  
We consider
the case where the light is resonant with the cavity, but the
detuning from atomic resonance is much larger than the 
natural atomic linewidth.
We thus analyze the coupled atom-cavity system after adiabatic 
elimination of the cavity and internal atomic dynamics
\cite{Doherty98}.
The resulting stochastic master equation (SME) in It\^o
form \cite{Kloeden92}  
for the atomic motion conditioned on the homodyne measurement,
describing the observer's best estimate for the atomic state, is 
\begin{equation}
  \setlength{\arraycolsep}{0ex}
  \renewcommand{\arraystretch}{1.5}
  \begin{array}{rcl}
    d\rho &{}={}& \displaystyle -i[\Heff,\rho]\, dt
      - \Gamma[\cos^2(\kt X), [\cos^2(\kt X), \rho]]\, dt \\
      &&\displaystyle -\sqrt{2\eta\Gamma}
          \{[\cos^2(\kt X), \rho]_+ - 2\langle\cos^2(\kt X)\rangle 
          \rho\}\, dW.
  \end{array}
  \label{adiabaticSME}
\end{equation}
The effective atomic Hamiltonian is
\begin{equation}
  \Heff = \pi P^2 - \Vmax\cos^2(\kt X),
\end{equation}
and $dW$ is given in terms of 
the scaled photodetector current (measurement record) $d\tilde{r}$
by
\begin{equation}
    dW = d\tilde{r}(t)/\sqrt{\eta} +
      \sqrt{8\eta\Gamma} \langle 
      \cos^2(\kt X)\rangle dt  .
   \label{mradiab}
\end{equation}
The physical interpretation of Eqs.~(\ref{adiabaticSME}) and (\ref{mradiab})
is that the photocurrent is a direct measurement of $\cos^2(\kt X)$
within the adiabatic approximation.
In these expressions, 
we are using scaled units such that position is measured in
units of $\sqrt{\hbar/(m\omegaHO)}$, momentum is measured in units
of $\sqrt{\hbar m\omegaHO}$, and time is measured in units of
$2\pi/\omegaHO$, where $\omegaHO:= \alpha g k \sqrt{2\hbar/|m\Delta|}$ 
is the oscillation frequency
in the harmonic approximation for one of the optical potential wells,
$g$ is the CQED coupling constant, 
$m$ is the atomic mass, 
$\Delta$ is the detuning from the atomic resonance, 
$\eta$ is the photodetection efficiency,
$\alpha$ is the mean cavity photon number in the absence of the
atom, 
$k$ is the optical spatial frequency, 
$\kt := k\sqrt{\hbar/(m\omegaHO)}$,
$\Vmax := \pi/\kt^2$, 
$\Gamma := {2\alpha^2 g^4}/{(\Delta^2\kappa\omegaHO)}$
is the effective measurement strength, 
$\kappa$ is the energy decay rate of the cavity, and
$dr(t) = \sqrt{\kappa/\omegaHO}\, d\tilde{r}(t)$ is the physical photocurrent.
The model thus has only three independent parameters, $\kt$, $\Gamma$, and 
$\eta$, in addition to any parameters that specify how the optical
potential is modulated.  
For the atomic parameters, we consider
cesium driven on the D$_2$ line, so that
$m=2.21\times 10^{-25}\;\mathrm{kg}$ and
$k=11732\;\mathrm{cm}^{-1}$.
For the cavity parameters, 
we choose values similar to those used in recent 
CQED experiments \cite{Hood98, Mabuchi99, Hood00},
yielding $\kappa/2\pi=40$~MHz, $g/2\pi = 120$~MHz, and $\alpha=1$.
We also assume an optical detuning of $\Delta/2\pi = 4\;\mathrm{GHz}$
to the red of the atomic resonance.
The corresponding scaled parameters are
$\Gamma=23.6$ and $\kt = 0.155$.  The scaled depth of the
optical potential is $\Vmax = 131$, 
so that the lowest two band energies
are within 1\% of their values in the harmonic approximation.
We also choose the idealized detection efficiency of $\eta=1$.
However, cooling performance similar to that shown below
is achievable even in the presence of detector inefficiency
($\eta$ well below $0.5$) as well as
other real-world limitations such as hardware propagation and
computation delays \cite{inprep}.

\begin{figure}[tb]
  \begin{center}
  \includegraphics[scale=0.5]{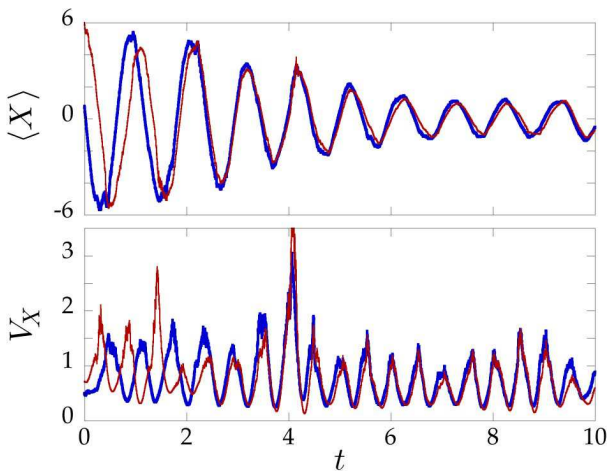}
  \end{center}
  \vspace{-5mm}
  \caption
        {Tracking behavior of the Gaussian estimator
         for the position centroid and variance for a sample
         trajectory.  
         The expectation values for the true wave packet (heavy
         lines) are shown with the corresponding Gaussian estimator
         quantities (narrow lines).  The cooling algorithm
         was activated at $t=2$. 
	\label{fig:basictracking}}
\end{figure}

While integration of the SME in order to obtain a state estimate 
is sensible in principle, in practice it is 
much too difficult to integrate the SME in real time.  
To solve this problem we must simplify the
estimator.  We make a Gaussian ansatz for the atomic
Wigner function (which, equivalently, prescribes how
the Weyl-ordered moments of the atomic state factorize).
In doing so, we reduce the effort of integrating a partial differential
equation to the much more manageable effort of integrating
a set of five coupled ordinary differential equations for the
two means $\langle X\rangle$ and $\langle P\rangle$ and the
three variances $V_x$, $V_p$, and $C_{xp}:=\langle XP+PX\rangle - 
\langle X\rangle\langle P\rangle$ \cite{EPAPS}.
The tracking behavior of the Gaussian estimator is illustrated
in Fig.~\ref{fig:basictracking}, which compares the evolution
of the position mean and variance for an atom evolving
according to Eq.~(\ref{adiabaticSME}) with the same quantities
from a Gaussian estimator evolution beginning
with a different initial condition.
The estimator rapidly locks on to the true solution.
The ability of the estimator to track both quantities well is
crucial to the success of the control algorithm that we present
here.

\begin{figure}[tb]
  \begin{center}
  \includegraphics[scale=0.45]{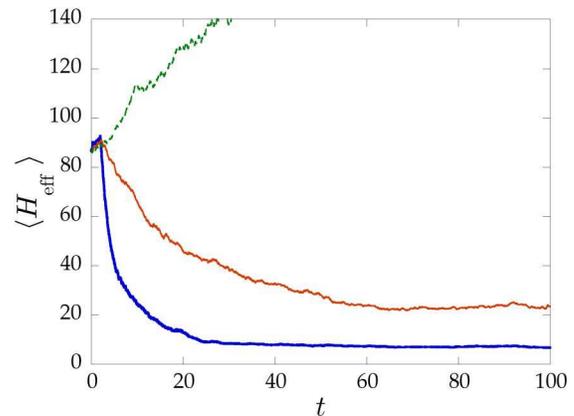}
  \end{center}
  \vspace{-5mm}
  \caption
        {Evolution of the mean effective energy 
         $\langle H_\mathrm{eff}\rangle$ (relative to the
         minimum potential energy), illustrating the 
         cooling-algorithm performance.
         Shown are: cooling via the Gaussian estimator
         (heavy solid), cooling based on the 
         photocurrent signal (narrow solid), and free
         atomic evolution with no cooling (dashed).
         Each curve is an
         average of 128 trajectories.
	\label{fig:basiccooling}}
\end{figure}

Intuitively, an algorithm that responds to the wave-packet centroid,
raising the optical potential while the atom is climbing
one side of a well, should cool the atom.  However, such a strategy
eventually breaks down due to uncontrolled pumping of energy into
the wave-packet variances.
The key to obtaining an effective control algorithm is to consider
the energy evolution due to the Hamiltonian part of the SME:
  $\partial_t\langle {H}_\mathrm{eff}\rangle_\mathrm{fb}
    = -(\partial_t \Vmax)\langle\cos^2(\kt X)\rangle$.
From this expression, it is clear that 
a ``bang-bang'' strategy is optimal for cooling if
we consider only cyclic modulations
of the potential amplitude within a limited range.
The potential should be switched to the extreme low value when
the quantity $-\langle\cos^2(\kt X)\rangle$ is maximized and to
the extreme high value when this quantity is minimized.
We denote these extreme values of the potential by
$(1-\varepsilon)^2\Vmax$ and $(1+\varepsilon)^2\Vmax$.
Such a control strategy is sensitive to energy in both the centroids
and variances of the atom.
Of course, it is not possible to tell whether
$-\langle\cos^2(\kt X)\rangle$ is at an extremum based only on 
information at the current time.  Thus, to complete the feedback
algorithm, we 
fit a quadratic curve to the history of this quantity at each time 
step and trigger on the
slope of this curve at the current time. 

\begin{figure}[tb]
  \begin{center}
  \includegraphics[scale=0.45]{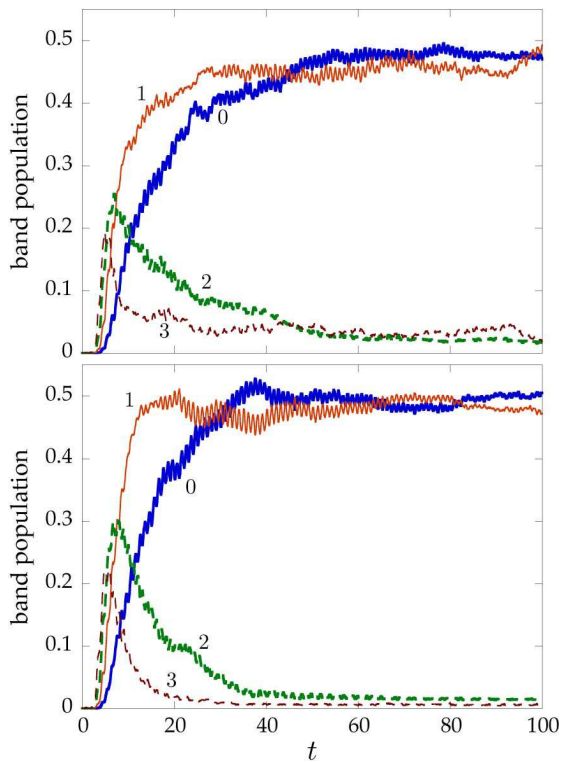}
  \end{center}
  \vspace{-5mm}
  \caption
        {Evolution of the band populations for the lowest
         four energy bands
         of the optical lattice, averaged over 128 trajectories.
         Top: cooling based on the Gaussian estimator. Bottom:
         cooling based on perfect knowledge of the actual
         wave function. 
	\label{fig:bandpops}}
\end{figure}

An example of the cooling algorithm in action is shown in 
Fig.~\ref{fig:basiccooling}, which compares the energy evolution
for an ensemble of atoms (where each atom is cooled separately
in an independent simulation) undergoing cooling
for $\varepsilon=0.1$ with 
the energy evolution without any cooling (i.e., free evolution
in the optical potential).  
Without cooling, the atoms simply
heat up due to the measurement back-action (or equivalently, the
stochastic cavity decay process),
until they are eventually lost from the optical potential.
The heating rate is set by the second term in the SME:
  $\partial_t\langle {H}_\mathrm{eff}\rangle_\mathrm{meas}
    = 8\pi\Gamma \langle \cos^2(\kt X)\sin^2(\kt X)\rangle$.
The cooling algorithm is able to counteract the measurement heating, 
and the atoms settle to a steady-state energy
much lower than the initial energy.  
Also shown in Fig.~\ref{fig:basiccooling} is the cooling performance
of a much simpler algorithm, which arises by noting that 
the photocurrent is a direct measure of 
$-\langle\cos^2(\kt X)\rangle$ (plus noise), and that it should be
possible to cool the atoms by triggering on the photocurrent signal
itself (an approach similar in spirit to the ``differentiating feedback''
strategy in Ref.~\cite{Fischer02}).  In this case, the same
quadratic curve-fitting procedure is used to temper the measurement
noise.  Although cooling to a steady state still occurs in this case,
the cooling performance is substantially worse than for the Gaussian
estimator.  This indicates that the Gaussian estimator acts as a 
nearly optimal filter (in the same sense as a Kalman filter
\cite{Doherty99}) for the relevant information in the 
photodetector signal.
The results shown were generated assuming initial conditions and 
curve-fit details
as follows, although the results do not sensitively depend on these 
values:
The atoms begin in a coherent state ($V_x = V_p = 1/2$),
and each has an initial centroid energy corresponding to a 
displacement from the well center
of 58\% of the distance to the edge of the well, chosen so the
atoms are clearly trapped but not particularly cold.
The initial locations of the atoms are distributed uniformly 
between the bottom of the well and this maximum distance.
The Gaussian estimator begins with the (impure-state)
initial conditions $\Xe = 6$, $\Pe = 0$, $\Vxe = \Vpe = 1/\sqrt{2}$,
and $\Ce = 0$. The estimator is evolved in steps of
$\Delta t=0.0005$, and at each step 
the quadratic curve is fitted to the
values of $-\langle\cos^2(\kt X)\rangle$ computed from the
300 most recent estimator steps.

The cooling performance is further illustrated in 
Fig.~\ref{fig:bandpops}, which shows the populations in the lowest
four bands of the optical potential.
Most of the steady-state population is in the lowest two bands,
accounting for 94\% of the population, and these two bands
are almost equally populated.  
We now show that this behavior can be understood
in terms of the parity of the atomic state.  The atomic
parity is invariant under Hamiltonian
evolution in the atomic potential (and hence the influence of
the control algorithm).
However, the measurement term in the SME causes
the atomic parity to diffuse according to 
\begin{equation}
  d\langle\prty\rangle = -\sqrt{8\eta\Gamma}
    \left[\langle\prty\cos(2\kt X)\rangle - \langle\prty\rangle
      \langle\cos(2\kt X)\rangle\right]\,dW ,
  \label{dparity}
\end{equation}
where $\prty$ is the parity operator.  In fact, this is an 
unbiased, nonstationary diffusion process that causes 
$\langle\prty\rangle$ to ``purify'' at late times to the
extreme values $\pm 1$ of pure parity,
so that the homodyne measurement acts like a QND measurement
of the atomic parity.
Since the atomic initial
condition described above corresponds to $\langle\prty\rangle=0$,
the measurement process drives the atoms to states 
with either pure even or pure odd parity with equal probability.
The obvious consequence of this effect 
for cooling is that only half of the atoms
can be cooled to the ground band, while the remaining half can
be cooled at best to the first excited band \cite{EPAPS}.  
However, due to this
purification effect (which means that the observer knows the final
state of the atom on each individual run), 
cooling to the first excited band is essentially
equivalent to cooling to the ground state, since the atom can be 
driven between these two bands with an additional coherent process
(e.g., modulation via an additional classical optical potential).

Much of the residual energy (excitation out of the lowest two
bands) is due to the inability of the Gaussian estimator to perfectly
track the true atomic state.  
In this light, the cooling performance is quite impressive, especially
considering the highly non-Gaussian 
nature of the first excited state, toward
which half the atoms are cooling.
To show this behavior explicitly, we can remove the estimator
from the problem, and repeat the same ensemble of simulations where 
the cooling algorithm operates based on the value of 
$-\langle\cos^2(\kt X)\rangle$ computed with respect to the
actual atomic wave function.  These results are also shown in 
Fig.~\ref{fig:bandpops}, and in this case the lowest two bands
account for 98\% of the atomic population.

\begin{figure}[tb]
  \begin{center}
  \includegraphics[scale=0.45]{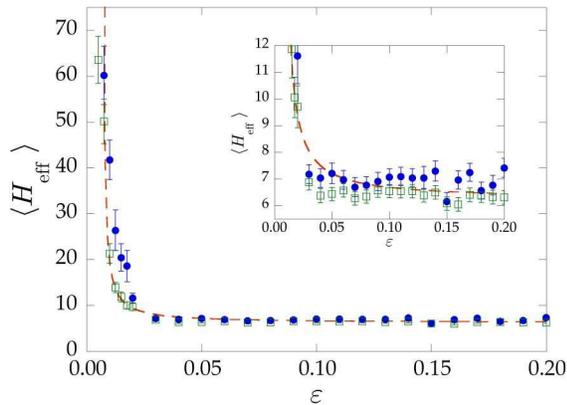}
  \end{center}
  \vspace{-5mm}
  \caption
        {Final energies, measured over the interval from $t=90$
         to $100$, as the switching amplitude $\varepsilon$
         varies.
         Circles: cooling based on the Gaussian estimator.
         Squares: cooling based on perfect knowledge of the 
         actual wave function.  Dashed line: simple cooling 
         theory, Eq.~(\ref{E0ss}).  Inset: magnified view of the
         same data. Error bars reflect
         standard errors from averages over 128 trajectories.
	\label{fig:epsilonsweep}}
\end{figure}

The steady-state atomic energies can be well described by a simple
analytic theory, which we now derive.  We will assume that the
atomic energy is near the ground-state energy $E_0$, so that we can make the
harmonic approximation for the potential.  We will further assume
for the moment that all the energy $\Delta E$ in excess of $E_0$ is associated 
with the wave-packet centroid (i.e., the wave packet is not squeezed).
Then the quantity $-\Vmax\langle\cos^2(\kt X)\rangle$ exhibits 
oscillations of peak-to-peak magnitude $\Delta E$.
Assuming that the switching occurs at the extremal times, 
the energy-evolution expression above leads to a
coarse-grained cooling rate of 
$\partial_t\langle {H}_\mathrm{eff}\rangle_\mathrm{fb}
= -8\varepsilon(\langle {H}_\mathrm{eff}\rangle - E_0)$.
On the other hand, the heating rate discussed above can be rewritten as
$\partial_t\langle {H}_\mathrm{eff}\rangle_\mathrm{meas}
= 4\Gamma\kt^4\langle {H}_\mathrm{eff}\rangle$.
Steady state occurs when these two rates sum to zero:
\begin{equation}
  \langle {H}_\mathrm{eff}\rangle_\mathrm{ss}
    = \frac{E_0+E_1}{2(1-\beta)}.
   \label{E0ss}
\end{equation}
Here, $\beta := \Gamma\kt^4/2\varepsilon$, and we have made the
replacement $E_0\rightarrow(E_0+E_1)/2$ in view of the above parity
considerations, where $E_1$ is the energy of the first excited band.

We can also derive a theory for the other extreme, where we assume
all the excess energy  is associated purely with 
squeezing of the wave packet and not with the centroid.
In this case, the oscillations of
$-\Vmax\langle\cos^2(\kt X)\rangle$
are of peak-to-peak magnitude
$\sqrt{\langle H_\mathrm{eff}\rangle^2-E_0^2}$.
Then the $(1-\beta)^{-1}$ dependence in 
Eq.~(\ref{E0ss}) is replaced by
$(1-\beta^2)^{-1/2}$.  In the regime $\beta \ll 1$ where this
theory should be valid, this prediction is always below that
of Eq.~(\ref{E0ss}), and thus the centroid-only theory more 
faithfully represents the physical cooling limit.

Eq.~(\ref{E0ss}) is compared with simulated late-time
energies in Fig.~\ref{fig:epsilonsweep}.  The simulated energies are
in good agreement with the simple theory, although they are 
generally higher than the simple theoretical prediction.  Again, much of 
this discrepancy is due to the imperfect tracking behavior of the 
Gaussian estimator, and the simulated energies from cooling based
on perfect knowledge of the actual wave function are in 
better agreement.  Qualitatively, we see a transition between 
ineffective and effective cooling with $\varepsilon$; essentially, 
$\varepsilon$ must be large enough that the cooling rate exceeds the
heating rate, ensuring that the system is controllable.
Beyond this transition, the system displays a remarkable insensitivity
to the value of the switching amplitude; we observe similar robustness
to the changes in the other parameters $\Gamma$, $\kt$, and $\eta$.


The authors would like to thank Andrew Doherty 
and Sze Tan for helpful discussions. This research was performed in 
part using the resources of the Advanced Computing Laboratory, 
Institutional Computing Initiative, and LDRD program
of Los Alamos National Laboratory.


\begin{thebibliography}{99}

\bibitem{Metcalf99}
H. J. Metcalf and P. van der Straten, \textit{Laser Cooling and
Trapping} (Springer, New York, 1999).

\bibitem{Raizen98}
M. G. Raizen et al.,
\pra \textbf{58}, 4757 (1998).

\bibitem{Belavkin99} 
V. P. Belavkin, 
Rep.\ Math.\ Phys., \textbf{43}, 405 (1999).

\bibitem{Doherty00} 
A. C. Doherty et al.,
\pra \textbf{62}, 012105 (2000).

\bibitem{Armen02}
M. A. Armen et al.,
\prl \textbf{89}, 133602 (2002).

\bibitem{Vitali03}
D. Vitali, S. Mancini, L. Ribichini, and P. Tombesi,
\josab \textbf{20}, 1054 (2003).

\bibitem{Hopkins03}
A. Hopkins, K. Jacobs, S. Habib, and K. Schwab,
\prb \textbf{80}, 235328 (2003).

\bibitem{Geremia03}
JM Geremia, J. K. Stockton, and H. Mabuchi,
arXiv.org preprint quant-ph/0309034.

\bibitem{Hood98} 
C. J. Hood, M. S. Chapman, T. W. Lynn, and H. J. Kimble, 
\prl \textbf{80}, 4157 (1998).

\bibitem{Mabuchi99} 
H. Mabuchi, J. Ye, and H. J. Kimble, 
Appl.\ Phys.\ B \textbf{68}, 1095 (1999).

\bibitem{Hood00} 
C. J. Hood et al.,
Science \textbf{287}, 1447 (2000).

\bibitem{Pinkse00} 
P. W. H. Pinkse, T. Fischer, P. Maunz, and G. Rempe, 
Nature (London) \textbf{404}, 365 (2000).

\bibitem{Fischer02} 
T. Fischer et al.,
\prl \textbf{88}, 163002 (2002).

\bibitem{Horak97} 
P. Horak et al.,
\prl \textbf{79}, 4974 (1997).

\bibitem{Wiseman02} 
H. M. Wiseman, S. Mancini, and J. Wang,
\pra \textbf{66}, 013807 (2002).

\bibitem{Wiseman93} 
H. M. Wiseman and G. J. Milburn, 
\pra \textbf{47}, 642 (1993).

\bibitem{Quadt95} 
R. Quadt, M. Collett, and D. F. Walls, 
\prl \textbf{74}, 351 (1995).

\bibitem{Doherty98} 
A. C. Doherty, A. S. Parkins, S. M. Tan, and D. F. Walls, 
\pra \textbf{57}, 4804 (1998).

\bibitem{Kloeden92}
P. E. Kloeden and E. Platen, \textit{Numerical Solution of 
Stochastic Differential Equations} (Springer, Berlin, 1992).

\bibitem{inprep}
D. A. Steck et al., in preparation.

\bibitem{EPAPS}
See EPAPS Document No.\ E-PRLTAO-92-053419 for sample animations
of two quantum trajectories and details regarding the Gaussian estimator.
A direct link to this document may be found in the online reference
section for this article at 
\href{http://link.aps.org/abstract/PRL/v92/e223004}{http://link.aps.org/abstract/PRL/v92/e223004}.

\bibitem{Doherty99} 
A. C. Doherty and K. Jacobs, 
\pra \textbf{60}, 2700 (1999).

\end{thebibliography}
\end{document}